\documentclass[]{interact}

\usepackage{lmodern}

\usepackage{epstopdf}
\usepackage{subfigure}

\usepackage[numbers,sort&compress,merge]{natbib}
\bibpunct[, ]{(}{)}{,}{n}{,}{,}
\renewcommand\bibfont{\fontsize{10}{12}\selectfont}
\renewcommand\citenumfont[1]{\textit{#1}}
\renewcommand\bibnumfmt[1]{(#1)}

\theoremstyle{plain}

\theoremstyle{definition}

\theoremstyle{remark}

\begin{document}

\title{Centrifugal Confinement Fusion Thruster}

\author{
  \name{Yi-Min Huang \thanks{CONTACT Yi-Min Huang Email: yopology@umd.edu}}
\affil{ Department of Astronomy, University of Maryland, College Park, Maryland
20742, USA}
}

\maketitle

\begin{abstract}
Centrifugal confinement fusion, a promising alternative to toroidal
confinement devices like tokamaks and stellarators, leverages supersonic
plasma rotation within a magnetic mirror configuration to achieve
simplified coil design, compactness, and enhanced stability. This
brief article explores the potential of centrifugal confinement fusion
for propulsion applications. A previous concern regarding the escape
of energetic ions, essential for propulsion, is addressed through
test-particle simulations. The results indicate that the earlier estimate
based on adiabatic invariance was overly pessimistic, and the underlying
physics is clarified.  

\end{abstract}


\section{Introduction}
Centrifugal confinement, an approach for magnetic confinement fusion,
utilizes supersonic plasma rotation to overcome the loss cone problem
and flute interchange mode, two major obstacles of a simple mirror
configuration\citep{Hassam1997,EllisHMO2001}. The centrifugal force
from the supersonic plasma rotation provides an effective gravitational
potential well that closes the ion loss cone except for the energetic
ions in the high-energy tail, and the associated velocity shear of
the rotation stabilizes the flute interchange modes\citep{Hassam1992,Hassam1999,Hassam1999a,HuangH2001,HuangGH2005}.
The simplicity of coil design, compactness, and stability make centrifugal
confinement an attractive alternative to toroidal confinement devices
such as tokamaks and stellarators. 

\begin{figure}
\includegraphics[width=1\textwidth]{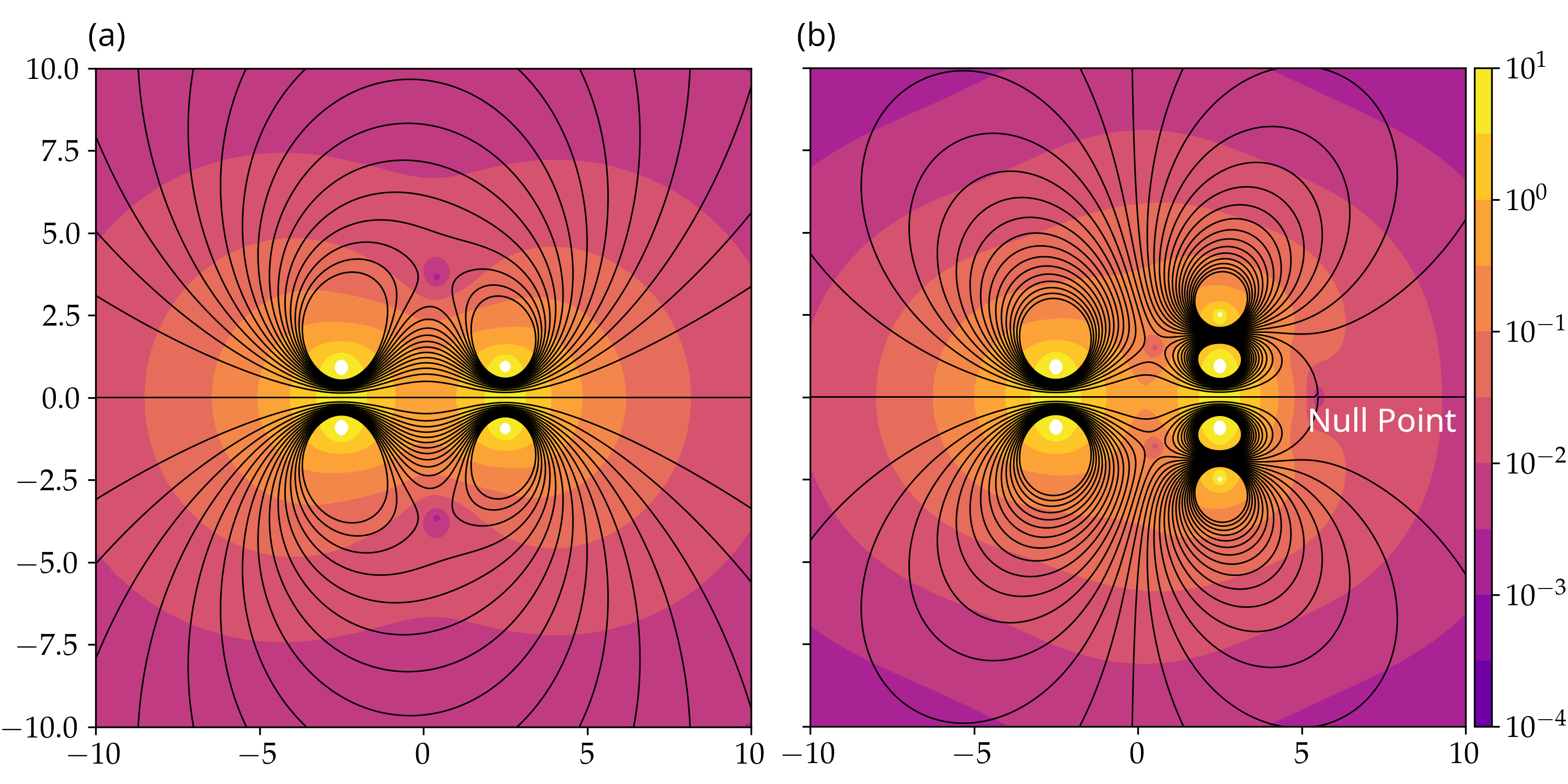}

\caption{(a) An asymmetric magnetic mirror machine with the mirror coil on
the right weaker than the one on the left. Energetic ions preferentially
exhaust from the weaker mirror, yielding a net propulsion to the left.
(b) An alternative asymmetric magnetic mirror machine with a third
coil with a dipole moment that exactly cancels the dipole moments
of the other two coils, making the global magnetic field is a quadrupole.
The magnetic field also has a null point on the right hand side of
the machine. Black lines are magnetic field lines, and color denotes
the magnitude of magnetic field in an arbitrary unit. \label{fig:asymmetric-magnetic-mirror}}
\end{figure}

Recently, a concept of utilizing a centrifugal confinement fusion
device for propulsion has been proposed. The idea is to make one of
the mirrors weaker such that the energetic ions preferentially escape
from the weaker mirror, thereby yielding a net propulsion (Figure~\ref{fig:asymmetric-magnetic-mirror}(a)).
For a reactor with one cubic meter of volume and a plasma density
of $10^{20}/\text{m}^{3}$, the propulsion force is estimated to be
500N, which can send a spacecraft of 1000kg from Earth to Jupiter
in about one month \citep{WhiteHB2018}.

However, a concern was raised regarding whether energetic ions can
actually escape from the mirror machine\citep{Hassam2018}.
The argument goes as follows. Ions and electrons are held together
by Coulomb interaction. Therefore, ions can escape only if electrons
can escape. Due to their lighter mass, electrons are more tightly
tied to magnetic field lines than ions. If the electrons remain tied
to the field lines, electrons escaped from one end will circle around
and return to the other end. If the electrons cannot escape, the energetic
ions will be constrained, preventing net propulsion. One may estimate
at what distance the electrons can break free from the field line
via the adiabatic invariance of the magnetic moment $\mu=mv_{\perp}^{2}/2B$.
Let $r$ be the distance from the mirror machine. A configuration
like Figure~\ref{fig:asymmetric-magnetic-mirror}(a) is a dipole
field at large scales, for which the magnetic field decreases with
the distance $r$ as $B\propto1/r^{3}$. Therefore, the perpendicular
speed $v_{\perp}\propto r^{-3/2}$ and the electron Larmor radius
$\rho_{e}\propto v_{\perp}/B\propto r^{3/2}.$ The adiabatic invariance
of $\mu$ requires $\rho_{e}\left|\nabla B\right|\ll B$. Conservatively,
the distance where the electrons become demagnetized may be estimated
by the condition $\rho_{e}\left|\nabla B\right|\sim B$. Let $\rho_{e0}$
be the electron Larmor radius inside the mirror machine and $a$ be
the characteristic size of the machine. From $\rho_{e}\sim\rho_{e0}\left(r/a\right)^{3/2}$
and $B/\left|\nabla B\right|\sim r$, the condition $\rho_{e}\left|\nabla B\right|\sim B$
yields $r/a\sim\left(a/\rho_{e0}\right)^{2}$. For a reactor with
the temperature $\sim10\text{keV}$ and the size $a\sim1\text{m}$,
the ratio $a/\rho_{e0}\sim10^{4}$; therefore, the electrons are demagnetized
at $r/a\sim10^{8}$, which is enormous. Although this estimate is
conservative (e.g., the requirement of $\rho_{e}\sim B/\left|\nabla B\right|$
may be excessive and collisions are neglected) and the electrons may
demagnetize at a shorter distance, the general idea suggests that
ions may be constrained by the electrons, limiting net propulsion.

To overcome this problem, an alternative magnetic coil configuration,
shown in Figure \ref{fig:asymmetric-magnetic-mirror}(b), has been
proposed \citep{Huang2018}.
In this new configuration, the original two mirror coils are of the
same strength, but a third coil of a larger radius is added next to 
the coil on the right. The electric current in the third coil
is opposite to the current in the first two coils, making the mirror
on the right weaker. Moreover, the current in the third coil is adjusted
to cancel the net dipole moment of the first two coils, making the
global magnetic field a quadrupole. Because the quadrupole field decreases
with $r$ as $B\propto1/r^{4}$, repeating the estimate for a quadrupole
field yields a condition for electron detachment at $r/a\sim a/\rho_{e0}\sim10^{4}$,
which is significantly less stringent than the previous estimate of
$10^{8}$ for a dipole field. An additional advantage of this new
configuration is the presence of a null point on the right. Because
electrons that get sufficiently close to the null point will be demagnetized,
the null point should also help the detachment of electrons.

\section{Test Particle Simulations}

\begin{figure}
\includegraphics[width=1\textwidth]{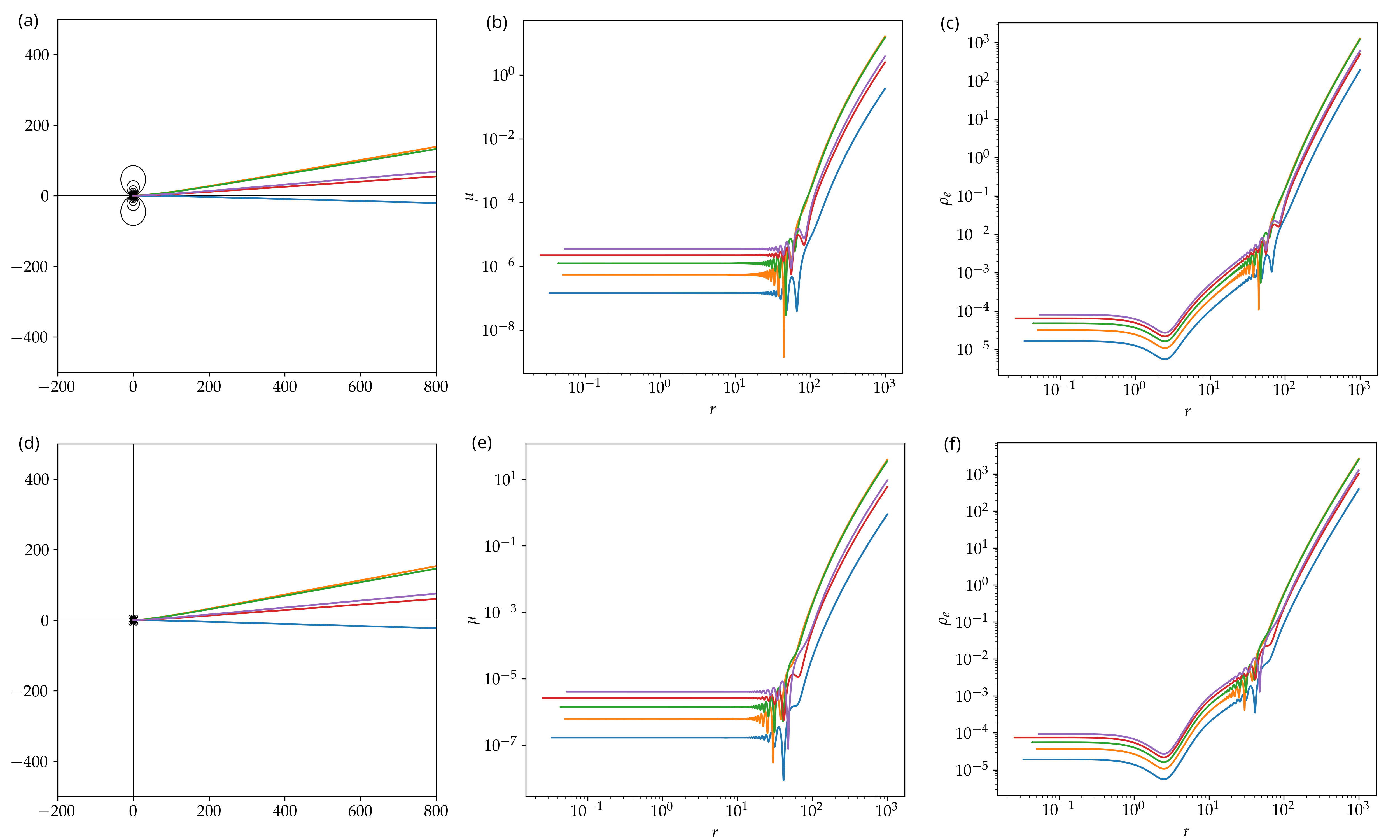}

\caption{(a) The trajectories, (b) the magnetic moments $\mu$ versus the distance
$r$, and (c) the Larmor radius $\rho_{e}$ versus $r$ for a sample
of electrons in a dipole magnetic mirror machine. (d) -- (f) The
results for a quadrupole magnetic mirror machine with electrons of
the same initial conditions.\label{fig:trajectories} }

\end{figure}

Given the estimates in the previous section, the key questions are:
Is it really very difficult for electrons to escape from a dipole
field as the estimate suggests? If that is the case, can the alternative
configuration of a quadrupole field come as a rescue? To address these
questions, test-particle simulations were performed in both dipole
and quadrupole fields with a Boris solver. For simplicity, only the
magnetic field is considered, and the electric field associated with
the plasma rotation is ignored. 

The simulations show that electrons readily escape the mirror machine
after passing the mirror throat, even in a dipole field. Figure \ref{fig:trajectories}
illustrates the trajectories, the magnetic moments $\mu$ with the
distance $r$, and the electron Larmor radius $\rho_{e}$ with the
distance $r$ for a sample of electrons in both dipole and quadrupole
configurations. These results indicate that the magnetic moment conservation
breaks down beyond $r\sim100$, and the electron Larmor radius $\rho_{e}$
becomes comparable to the distance $r$ from the mirror at $r\sim1000$.
Interestingly, there is no significant difference between the dipole
and the quadrupole field results, with the quadrupole field slightly
accelerating the breakdown of magnetic moment conservation. 

What causes the conservation of magnetic moment to break down earlier
than expected in a dipole field? Upon further reflection, it is realized
that the adiabatic invariance of $\mu$ requires not only $\rho_{e}\left|\nabla B\right|\ll B$
but also $dB/dt\ll\omega_{ce}B$, where $\omega_{ce}$ is the electron
gyro-frequency. These two conditions are typically comparable if the
parallel speed $v_{\parallel}$ and the perpendicular speed $v_{\perp}$
are comparable. In the present case, however, as electrons move away
from the magnetic mirror, the perpendicular speed decreases as $v_{\perp}\propto r^{-3/2}$,
whereas $v_{\parallel}$ increases to conserve the total energy. Therefore,
the condition $dB/dt\ll\omega_{ce}B$ must be considered separately. 

Qualitatively, we may estimate $dB/dt\sim v_{\parallel}|\nabla B|$,
where $v_{\parallel}$ is on the order of the initial electron thermal
speed $v_{0}$. For a dipole field, the gyro-frequency $\omega_{ce}\sim v_{\perp}/\rho_{e}\sim(a/r)^{3}v_{0}/\rho_{e0}$.
Hence, the condition $dB/dt\ll\omega_{ce}B$ yields $r/a\ll\left(a/\rho_{e0}\right)^{1/2}$.
This is a much weaker condition than the previous estimate of $r/a\ll\left(a/\rho_{e0}\right)^{2}$
for a dipole field. The physical picture goes as follows. As electrons
move away from the mirror, their gyro-period lengthens, allowing them
to traverse significant distances within a single gyro-period. This
leads to a significant change in the magnetic field within a gyro-period,
causing the breakdown of the magnetic moment conservation. For a fusion
reactor with $a/\rho_{e0}\sim10^{4}$, this breakdown occurs at $r/a\sim100$,
aligning with the simulation results.

\section{Conclusion}

In conclusion, this study shows that electrons are not as strongly
attached to the magnetic mirror dipole field as suggested by an earlier
pessimistic estimate. The gap in the previous estimate regarding the
escape condition has now been addressed and clarified. Consequently,
the concern about constraint on energetic ion escape due to electrons,
which could have limited net propulsion, is no longer a significant
issue. The results also indicate that a quadrupole field may not be
necessary, because a dipole field is already sufficient.

Centrifugal confinement, therefore, shows promise for both fusion
power generation and space propulsion. Built upon the success of the
Maryland Centrifugal Experiment (MCX) at the University of Maryland,
College Park, the Centrifugal Mirror Fusion Experiment (CMFX)\citep{Romero2021}
 at the University of Maryland, Baltimore County, has recently achieved
significant advancements. These include steady-state operation for
several seconds and the detection of neutrons from Deuterium-Deuterium
fusion\citep{Romero2024}. The future of centrifugal confinement appears bright, with exciting
developments expected from future generations of this technology.

Acknowledgments --- Beneficial discussion with Prof.~A.~B.~Hassam is gratefully acknowledged.

\bibliographystyle{tfp}

\bibliography{ref}

\end{document}